\input phyzzx
\hoffset=0.375in

\def\kms{\rm km\,s^{-1}}

\font\bigfont=cmr17
\centerline{\bigfont The Zero Point of Extinction Toward Baade's Window} 
\bigskip
\centerline{{\bf Andrew Gould}\foot{Alfred P.\ Sloan Foundation Fellow},
{\bf Piotr Popowski}, and \bf Donald M.\ Terndrup}
\bigskip
\centerline{Dept of Astronomy, Ohio State University, Columbus, OH 43210}
\smallskip
\centerline{E-mail:  gould,popowski,terndrup@astronomy.ohio-state.edu}
\bigskip
\singlespace
\centerline{\bf ABSTRACT}
	We measure the zero point of the Stanek (1996) extinction map
by comparing the observed $(V-K)$ colors of 206 K giant stars with their
intrinsic $(V-K)_0$ colors as derived from their H$\beta$ indices.  We find
that the zero point of the Stanek map should be changed by 
$\Delta A_V = -0.10\pm 0.06$ mag, obtaining as a bonus a three-fold
reduction of the previous statistical error.  The most direct way to test
for systematic errors in this determination would be to conduct a parallel
measurement based on the $(V-K)$ colors of RR Lyraes (type ab).

Subject Headings:  Galaxy: general -- Hertzsprung-Russell diagram
\endpage

\chapter{Introduction}

	Baade's Window, $(\ell,b)\sim (1^\circ,-4^\circ)$, has been 
an important
laboratory for the study of bulge populations. The key features
of Baade's Window that made it such a focus of early work were its relatively
low extinction $(A_V\sim 1.5)$ and the presence of NGC 6522 which 
provided an opportunity to measure that extinction.  The value of this window
increased significantly when Stanek (1996) constructed a detailed extinction
map with $30''$ resolution by applying the method of Wo\'zniak \& Stanek (1996)
to observations by OGLE (Szyma\'nski \& Udalski 1993; Udalski et al.\ 1993)
over a $40'$ square field.  It is now possible to deredden stellar samples
on almost a star-by-star basis in this field.  Stanek (1996) estimates
the error in {\it differential} extinction to be $\sim 0.1\,$mag in $A_V$,
but notes that the errors in the absolute extinction are dominated by the 
zero-point error, $0.20\,$mag.  
For many applications, such as the interpretation
of color-magnitude diagrams of bulge field stars and of the cluster NGC 6522
or the measurement of distances using RR Lyraes or other tracers, the 
determination of the zero point is crucial.

	Here we argue that the best way to estimate the zero point for the
$A_V$ map is to measure $\Delta E(V-K)_i$ for an ensemble of stars $i=1...n$,
defined by
$$\Delta E(V-K)_i = (V-K)_{i} - (V-K)_{0,i} - 
(1-\alpha) A_{V,i}^{\rm Stanek},\eqn\deltavmk$$
where $(V-K)_{i}$ is the observed color of the star, $(V-K)_{0,i}$
is its predicted unreddened color, $A_{V,i}^{\rm Stanek}$ is the 
visual extinction
at the position of the star in the Stanek (1996) map, and $\alpha$ is the
ratio of the extinction in the $K$ and $V$ bands, assumed to be 
(Rieke \& Lebofsky 1985),
$$\alpha \equiv {A_K\over A_V} = 0.11.\eqn\rkdef$$
The correction to the zero point of the Stanek (1996) map is then given
by 
$$\Delta A_V = {\langle \Delta E(V-K)\rangle\over 1-\alpha}, \eqn\deltaak$$
where $\langle \Delta E(V-K)\rangle$ is a suitably weighted average of
equation \deltavmk\ over the sample.
	We then apply this method to the data of Terndrup, Sadler, \& Rich 
(1995, TSR) and find
$$\Delta A_V = -0.10\pm 0.06.\eqn\finalresult$$ 
That is,
the $A_V$ of stars in Baade's Window are on average $0.10\,$mag {\it lower}
than previously believed.  

\chapter{The Stanek Extinction Map}

	Stanek (1996) measured the mean position
of the red giant clump on the color-magnitude diagram ($\langle V\rangle$,
$\langle V-I\rangle$) as a function of position on the sky $(x,y)$.  He then
estimated the differential
total and differential selective extinctions,
$$A_V(x,y) =\langle V\rangle(x,y) + C_V,\qquad E{(V-I)}(x,y) 
= \langle V-I\rangle(x,y) + C_{V-I},
\eqn\avevmi$$
with the constants, $C_V$ and $C_{V-I}$, being undetermined.
He found empirically a very strong correlation between $A_V$ and $E{(V-I)}$:
$$ A_V(x,y) \simeq
 2.49\,E{(V-I)}(x,y)+ C. \eqn\corr$$
On physical grounds, $C=0$.  This left one constant, either $C_V$ or $C_{V-I}$,
to be determined.

	Stanek (1996) then determined the overall zero point using the 
measurement $E{(V-I})=0.59 \pm 0.08$ made by TSR in a sub-region of Baade's 
Window, Blanco region A (Blanco, McCarthy \& Blanco 1984).  That is, he
set the zero point so that the average of $E{(V-I)}$ over this region of his
map reproduced the TSR value.  He then used equation \corr\ (with $C=0$)
to establish the zero point of the $A_V$ map; 
i.e., he set the mean extinction of region A to
be $\langle A_V\rangle=0.59\times 2.49=1.47$.  The formal error in this
determination is therefore $0.08\times 2.49=0.20$.

\chapter{Previous Approaches}

	With the exception of TSR, all previous determinations of 
the extinction toward Baade's Window have been made by measuring the
selective extinction $E(B-V)$, and then multiplying by
an assumed ratio of total to selective extinction $R_V=A_V/E(B-V)$
(Arp 1965; van den Bergh 1971; Walker \& Mack 1986; Terndrup \& Walker 1994).
There are several major disadvantages to this approach.  First, $R_V\sim 3$
is rather large, and the statistical error in $E(B-V)$ (usually estimated
to be $\geq 0.03$) is multiplied by this factor when estimating the error
in $A_V$.  Second, $R_V$ varies along different lines of sight, so for any
particular line of sight for which it is not actually measured (e.g., Baade's
Window) the precision of the estimate is no better than 7\%.  Hence, the
statistical error alone for $A_V$ is more than 0.12 mag.  Finally, there
are systematic errors arising from uncertainties in the intrinsic $(B - V)$
colors of stars used to estimate $E(B-V)$.  While the intrinsic color of
extremely hot stars (in the Raleigh-Jeans limit) is known from fundamental 
physics, there are no such
stars lying beyond the dust column in this direction.  Hence, one must use
cooler stars whose $(B - V)$ colors are sensitive functions of temperature,
metallicity, and
perhaps other factors.  The standard approach is to find local analogs of
the program stars and directly measure their colors, but systematic errors
may arise from any unrecognized differences between these two groups of stars.
As always, it is difficult to determine the size of the systematic errors, but
one can gain a sense of their
magnitude by comparing the $E(B-V)_0=0.45\pm 0.04$ derived
by van den Bergh (1971) from three different methods based on cool stars
(K and M giants) with the $E(B-V)_0=0.60\pm 0.03$ derived by Walker \& Mack
(1986) using relatively hot stars (RRab Lyraes).  Here the subscript ``0'' 
means  ``reduced
to zero color'' using the prescription adopted by TSR from Dean, Warren, \&
Cousins (1978).

	TSR pioneered a radically different approach, although they did
not call explicit attention to this fact.  They measured 
$$E{(V-K)} = 1.23\pm 0.08,\eqn\evk$$
for the Blanco A region by comparing the $H\beta \, \lambda 4861$
(Faber et al. 1985) index as a function of calculated $(V-K)$ color
to that observed for bright K giants in the solar neighborhood.
They then inferred (but did not explicitly write down),
$$A_{V} = {E(V-K)\over 1-\alpha} = 1.38\pm 0.09.\eqn\ak$$
While this approach is formally identical to the previous one (measurement of
a selective extinction and conversion to a total extinction), it is 
potentially more accurate than using $E(B-V)$ because the extrapolation to 
total visual extinction is small (a factor 1.12 vs.\ 3),
and therefore the error in $A_V$
is only slightly bigger than the error in $E(V-K)$.
TSR then used this measurement to infer $E{(B-V)_0}=0.47\pm0.04$
and $E{(V-I)}=0.59\pm 0.08$.
To obtain these quantities, they had to employ estimates of the ratios of
total to selective extinctions which accounts for the proportionately larger
error bars compared to those in equations \evk\ and \ak.  In 
particular, TSR assumed
$$A_V = 2.33 E(V-I),\eqn\akt$$
where the coefficient is considerably lower than Stanek's (1996) empirical
value for Baade's Window [eq.\ \corr].  
By fixing the zero point according to TSR's $E(V-I)$
rather than TSR's more directly determined $A_V$, Stanek (1996) therefore
overestimated the visual extinction by $(2.49-2.33)\times 0.59= 0.09$ mag, 
and also
overestimated the uncertainty.  From this simple argument, we therefore
derive a naive correction to the Stanek (1996) extinction map,
$$\Delta A_V = -0.09 \pm 0.09 \qquad (\rm naive).\eqn\naive$$

\chapter{New Determination}

	However, rather than simply adopting the correction given by 
equation \naive, we prefer for several reasons to make a complete 
redetermination of the zero point of the Stanek (1996) extinction map.  
First, Stanek (1996) compared the TSR extinction
for the {\it whole} of region A with the mean extinction of the {\it subregion}
of region A that is covered by his map.  About 35\% of the TSR stars lie 
within $2'$ of NGC 6522 and so are excluded from the map.  One should therefore
compare the mean extinction of the TSR stars lying within the map with the
mean extinction predicted by the map for their positions, otherwise 
systematic trends of extinction with position could affect the result.  Second,
TSR measured and reported the mean extinctions in region A and in regions B 
and C separately.  They did so because these areas have different average
extinctions (Blanco et al.\ 1984), so
it would degrade the information content of their results to combine the
two.  However, for purposes of measuring the {\it offset} to the Stanek (1996)
map,
the fact that different regions have different extinctions is irrelevant.
The only concern is to measure the {\it difference} in the observed 
extinction at each point from that predicted by the map.  Including all three
regions approximately triples the sample and correspondingly reduces the
statistical errors.  Third, TSR used the relatively crudely determined
H$\beta$/$(V-K)_0$ relation of Faber et al.\ (1985). A much more sophisticated
polynomial relation is available from Gorgas et al.\ (1993, Table 6, eq.\ 5):
$$\eqalign{{\rm H}\beta = & 8.2261  - 5.9295(V-K)_0 + 0.52968(\log g) 
- 0.048352(\log g)^2 \cr & 
 - 0.23695(\log g)(V-K)_0 + 1.8169(V-K)_0^2 - 0.19721(V-K)_0^3,}\eqn\gorgas$$
where we have ignored the terms in [Fe/H] since they are small and since
$\langle\rm [Fe/H]\rangle\sim 0$ for the sample in any case.  

	To carry out our analysis, we first restrict the sample to the 209 
stars with
$$2.1 \leq (V-K)_* \leq 3.0,\eqn\vmkrange$$
where
$$(V-K)_*\equiv (V-K) - 
(1-\alpha)(A_{V}^{\rm Stanek}+\Delta A_{V,*}), \eqn\vmkstar$$
is our final best-fit estimate of $(V-K)_0$ as determined
from our final best-fit offset to the Stanek (1996) map [eq.\ \finalresult].
The upper limit is chosen to exclude M giants which have TiO bands that 
influence the $H\beta$ index. Gorgas et al.\ (1993) also excluded these stars
when they derived equation \gorgas.  We note, however, that if we extend
the sample to the 239 with $(V-K)_*<3.8$ (the range of the Gorgas et al.\
1993 data),
the results change by less than $0.01$ mag.  
The lower limit in equation \vmkrange\ is justified below.

	To obtain $(V-K)$ from the $V$ and $I$ measurements, we slightly modify
the procedure of TSR.  TSR converted from $(V-I)$ to $(V-K)$ colors  based
on the extremely tight quadratic color-color relation obtained for a patch
of Baade's Window by Tiede, Frogel, \& Terndrup (1995).  We modify this 
procedure by using the Stanek (1996) differential map to transform each $(V-I)$
from its observed value to the value it would have if it lay in the Tiede 
et al.\ (1995) region and then use the same relative extinction to transform 
the inferred $(V-K)$ back to the star's actual position.  In practice,
the individual corrections generated by this procedure are small, typically
$<0.01$ mag, because the color-color track is almost parallel to the reddening
vector.  The net effect on the final result is $\ll 0.01$ mag.

	We estimate the surface gravity $g$ from the inverse square of the
star's effective radius in $K$ band,
$$\log {g\over g_\odot} = \log{S\over S_\odot} + 
0.4(K - \alpha A_V - M_{K,\odot}- \mu),\eqn\logg$$
where $S\propto [\exp(hc/\lambda_K k T)-1]^{-1}$ is the black body
surface brightness at $K$ band ($\lambda_K=2.2\mu$m) for an assumed temperature
$T= 8520 - 2230(V-K)_0 + 267(V-K)_0^2$ obtained by fitting the values given
in Table 4 of Ridgway et al.\ (1980).  Here $\mu=14.5$ is the adopted distance 
modulus to the Galactic center and $M_{K,\odot}=3.3$.

	For each trial value of $\Delta A_V$, we deredden the ``observed'' 
$(V-K)$ 
color (inferred from $V-I$) and $K$ magnitude of each star, use these to
estimate its temperature and surface gravity, and finally predict H$\beta$
from equation \gorgas.  We form $\chi^2(\Delta A_V)$ from the difference of 
the observed H$\beta$ and these predictions divided by the errors as 
reported by TSR. The best-fit $\Delta A_V$ and its errors are determined 
from this function.

\FIG\one{
Observed H$\beta$ index ({\it solid squares}) compared to the values predicted
({\it open circles}) from the relation of Gorgas et al.\ (1993).  The
prediction is mainly a function of $(V-K)_*$, the best-fit estimate of
$(V-K)_0$, using the observed $(V-K)$ color, the Stanek (1996) extinction
for the position of each star, and the overall offset given by 
eq.\ \finalresult.  The points are binned by $\sim 23$ stars and the error
bars indicate the standard error of the mean based on the individual errors
reported by TSR.  The stars $(V-K)_*>3.0$ (right dashed line) were excluded
from the fit because they are M giants which are affected by TiO bands and
because Gorgas et al.\ (1993) excluded such stars when they
determined eq.\ \gorgas.  The stars $(V-K)_*<2.1$ (left dashed line) were 
excluded from the fit because they are contaminated with foreground stars
and hence fall systematically below the predicted values (see text).
}

	We find that in the adopted interval, $2.1\leq(V-K)_*\leq 3.0$, the
predicted and observed H$\beta$ indices (weighted by the observational errors)
are in good overall agreement with $\chi^2=179$ for 208
degrees of freedom.  However, the observations deviate markedly from the
predictions for $(V-K)_*< 2.1$ (see Fig.\ \one, below).  These bluer stars
are mostly G giants and subgiants.
The original sample from which TSR drew their stars was selected primarily
for proper motion studies and hence was composed of preferentially brighter
stars.  The intrinsically fainter G stars are therefore likely to be 
foreground disk stars.  From equations \gorgas\ and \logg, we find that
if one of these star lies $\Delta \mu$ in the foreground, H$\beta$ will be
overestimated by $\sim 0.1\Delta\mu$.  Thus, this selection effect can
explain at least some of the observed deviation.  In any event, these stars
clearly differ from the bulk of the K giants (which dominated the fit by
Gorgas et al.\ 1993) and we therefore restrict the sample to $(V-K)_*\geq 2.1$.

	We now consider various systematic effects.  First, there are several
assumptions that affect the results through the estimate of the surface
gravity.  For example, we find that if the adopted distance to the Galactic
center is increased by $\Delta\mu$, then the surface gravities are decreased,
implying that $\Delta A_V$ is also decreased (becomes more
negative) by $-0.11\Delta\mu$.  Similarly, if the mean mass of the stars
is larger than one solar mass (the value that
 we implicitly assumed in writing eq.\
\logg) then $\Delta A_V$ is increased by $0.10\Delta \ln M$.  If the
effective radius at which surface gravities are measured differs from that
of the K band photosphere, $\Delta A_V$ is changed by $-0.20\Delta \ln R$.
If the adopted temperatures are on average different from the true ones,
$\Delta A_V$ is changed by $0.19\Delta \ln T$.
Plausible adjustments for each of these factors are therefore likely to affect
$\Delta A_V$ by $\lsim 0.01\,$mag.  

	Gorgas et al.\ (1993) report a scatter in their fit to the H$\beta$ 
index of 0.28.  We estimate that this scatter causes an uncertainty in the
zero point of equation \gorgas\ of $\sim 0.025$, which translates directly
into an uncertainty in $\Delta A_V$ of 0.03 mag.

	Next, we break the sample into two subsamples, one of stars
in region A (where TSR found $A_V=1.38$) and one of stars in regions 
B and C (where they
found $A_V=1.55$).  We obtain separate fits of $\Delta A_V =-0.07\pm0.09$
and $\Delta A_V =-0.13\pm0.06$, which are consistent at the $1\,\sigma$
level.

	Finally, we divide the sample into two subsamples, according to their
angular distance $\theta$ from the center of NGC 6522.  The inner group 
$(2.\hskip-2pt ' 0<\theta<2.\hskip-2pt ' 5)$ has 49 stars and the outer group
$(\theta>2.\hskip-2pt ' 5)$ has 160 stars.  We find values of
$\Delta A_V = -0.28\pm 0.12$ and $\Delta A_V = -0.06\pm 0.06$, respectively.
That is, they are inconsistent at the $1.6\,\sigma$ level.  This is somewhat
worrisome because it may indicate that the inner group is affected in some
way by proximity to the cluster.  We investigate the following possible
effects.  First, Stanek (1996) reports that the extinction values close
to NGC 6522 and NGC 6528 are systematically lower than surrounding
regions, leading him to believe that he may have underestimated it (due to
some unspecified form of contamination).  This was the primary reason for
excluding the region $\theta<2'$ from the map.  
It is possible that this effect, if real,
extends beyond $2'$.  However, the {\it sign} of the effect is wrong to
explain the difference between the two subsamples.  It is
possible that contamination by cluster stars generates some other effect
that has the correct sign.  We perform the following tests to search for
cluster contaminants.  First we search for an excess of stars in the 
underlying sample (which reaches as close as $\theta \sim 1'$ from the cluster
center) with radial velocities 
that are consistent with the 
cluster velocity, $\sim -25\,\kms$ (Rich 1990; 
Smith, Hesser \& Shawl 1976), both in
the sample as a whole and as a function of $\theta$.  If contamination by the
cluster were significant, one would expect an excess, 
especially at small radii.  None is detected.
Next we conduct a test that is sensitive to even lower levels of
contamination: we plot the proper motions of all stars having radial
velocities consistent with cluster membership.  If even a small subset of these
are in the cluster, the proper-motion diagram should show a clump.  
We detect a common proper motion clump of eight stars.  We will report 
elsewhere on this measurement of the proper motion of NGC 6522.  For present
purposes, we note that three of the eight stars are part of our sample, all
three being in the inner annulus.  We exclude these, leaving a sample
of 206 stars.  The estimates for $\Delta A_V$ in the inner and outer annuli 
are then $-0.25\pm 0.12$ and $\Delta A_V = -0.06\pm 0.06$, respectively,
a $1.4\,\sigma$ difference.  Since contamination by the cluster has been
eliminated, this difference should be regarded as a normal statistical
fluctuation, and we therefore include the entire remaining sample of 206 stars
and find
$$\Delta A_V = -0.10\pm 0.05\qquad ({\rm internal}\ {\rm error)}
.\eqn\finalresultprime$$ 
We then add in quadrature
the external calibration error of 0.03 mag intrinsic to the
Gorgas et al.\ relation \gorgas\ to obtain equation \finalresult.  
Figure \one, shows the mean
predicted and observed H$\beta$ indices (weighted by the observational errors)
for the overall best fit, binned by $(V-K)_*$.  We find $\chi^2=176$ for 205
degrees of freedom.

For completeness we note that had we adopted the relation of Faber et al.\
(1985) in place of equation \gorgas\ from Gorgas et al.\ (1993), $\Delta A_V$
would decrease from $-0.10$ to $-0.11$.

\chapter{Future Tests}

	Of course, the most important potential source of systematic errors
is not probed by the tests of the previous section:
the possibility that the stars in Baade's Window differ systematically in
some unknown way from the local stars upon which equation \gorgas\ is based.
Ultimately, the only way to test for this effect is to make an independent
determination of $E(V-K)$ on a substantially different set of stars.
The obvious choice for this test is RR Lyraes (type ab).  First, the comparison
between RRab's and cool giants showed the largest discrepancy of all
determinations based on $E(B-V)$, so it is important to see if this discrepancy
persists for $E(V-K)$.  Second, because of the relatively narrow range of RRab 
colors and the accuracy of the Stanek (1996) map, we estimate that each star 
should provide
a statistically independent estimate of the zero point accurate to 0.15 
magnitudes.  Since there are more than 50 such stars in the region, the limit
for this method is set by the size of the calibrating sample (17) observed
by Jones et al.\ (1992).  We estimate this limiting uncertainty to be only
$\pm 0.03$ mag.  Work is in progress to apply this method. 

{\bf Acknowledgements}:  
Work by AG and PP was supported in part by grant AST 94-20746 from the NSF.
Work by DMT was supported in part by grant AST 95-28227 from the NSF.

\endpage
\Ref\arp{Arp, H.\ 1965, ApJ, 141, 43}
\Ref\bla{Blanco, V.\ M., McCarthy, M.\ F., \& Blanco, B.\ M. 1984, AJ, 89, 636}
\Ref\dwc{Dean, J.\ F., Warren, P.\ R., \& Cousins, A.\ W.\ J.\ 1978, MNRAS, 
183, 569}
\Ref\faber{Faber, S. M., Friel, E. D., Burstein, D., \& Gaskell, C.\ M.\ 1985
ApJS, 57, 711}
\Ref\gor{Gorgas, J., Faber, S. M., Burstein, D., Gonz\'alez, J., Courteau, S.,
\& Prosser, C.\ 1993, ApJS, 86, 153}
\Ref\john{Johnson, H., L.\ 1965, ARA\&A, 4, 193}
\Ref\jones{Jones, R.\ V., Carney, B.\ W., Storm, J., \& Latham, D.\ W.\ 1992,
ApJ, 386, 646}
\Ref\ric{Rich, R.\ M.\ 1990, ApJ, 362, 604}
\Ref\ridg{Ridgway, S.\ T., Joyce, R.\ R., White, N.\ M. \& Wing, R.\ F. 1980,
ApJ, 235, 126}
\Ref\rieke{Rieke, G.\ H., Lebofsky, M.\ J. 1985, ApJ, 288, 618}
\Ref\shs{Smith, M.\ G., Hesser, J.\ E., \& Shawl, S.\ J.\ 1976, ApJ, 206, 66}
\Ref\stan{Stanek, K.\ Z.\ 1996, ApJ, 460, L37}
\Ref\su{Szyma\'nski, M., \& Udalski, A.\ 1993, Acta Astron., 43, 91}
\Ref\tsr{Terndrup, D.\ M., Sadler, E.\ M., \& Rich, R.\ M.\ 1995, AJ, 110, 
1774 (TSR)}
\Ref\tw{Terndrup, D.\ M., \& Walker, A.\ R.\ 1994, AJ, 107, 1786}
\Ref\tft{Tiede, G.\ P., Frogel, J.\ A., \& Terndrup 1995, AJ, 110, 2788}
\Ref\ud{Udalski, A.\ et al. 1993, Acta Astron., 44, 165}
\Ref\vdb{van den Bergh, S.\ 1971, AJ, 76, 1082}
\Ref\wm{Walker, A. R., \& Mack, P.\ 1986, MNRAS, 220, 69}
\Ref\ws{Wo\'zniak, P.\ R., \& Stanek, K.\ Z.\ 1996, ApJ, 464, 233}

\endpage
\refout
\figout
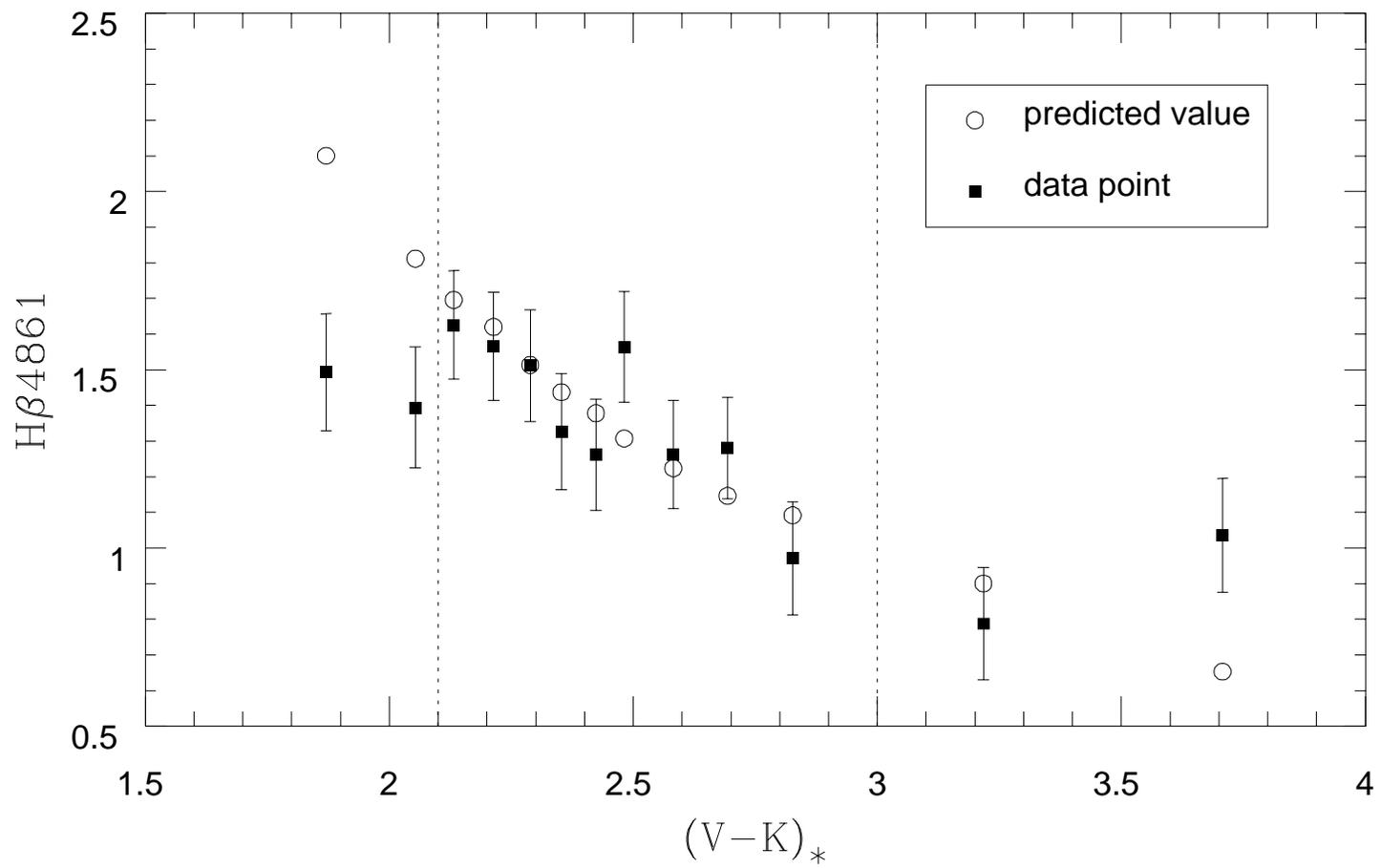
\end